\documentclass{ws-rv9x6}
\usepackage{epstopdf}

\def\figDIR{./}

\begin{document}
\setcounter{chapter}{8}
\chapter{Auxiliary-field quantum Monte Carlo methods in nuclei}\label{alhassid}

\author[Y. Alhassid]{Y.  Alhassid\footnote{yoram.alhassid@yale.edu} } 

\address{Center for Theoretical Physics, Sloane Physics 
Laboratory,\\ Yale University, New Haven, Connecticut 06520, USA}

\begin{abstract}
Auxiliary-field quantum Monte Carlo methods enable the calculation of thermal and ground state properties of correlated quantum many-body systems in model spaces that are many orders of magnitude larger than those that can be treated by conventional diagonalization methods. We review recent developments and applications of these methods in nuclei using the framework of the configuration-interaction shell model.
\end{abstract}

\body 
\newpage
\section{Introduction}

A major challenge in nuclear theory is understanding the properties of nuclei from the underlying interactions between their constituents.  There has been much progress in applying {\it ab initio} methods such as the Green's function Monte Carlo~\cite{GFMC,GFMCa}, the no-core shell model~\cite{Barrett20013,Stetcu2013}, and the symmetry-adapted no-core shell model~\cite{Draayer2012}  to calculate nuclear properties,  but these approaches are limited to light nuclei. The coupled-cluster method~\cite{Hagen2014} has been applied to light nuclei and mid-mass nuclei near shell closure. Density functional theory~\cite{Bender2005,Bertsch2007} is applicable across the table of nuclei, but as a mean-field approximation it can miss important correlations and the description of excited states requires  extensions of the theory. 

The configuration-interaction (CI) shell-model approach~\cite{Caurier2005} is a suitable framework to include correlations beyond the mean-field approximation. The CI shell model is widely used in nuclear, atomic and molecular physics. However, the dimensionality of the many-particle model space scales combinatorially with the number of valence single-particle orbitals and the number of valence nucleons, hindering its applications in mid-mass and heavy nuclei. 

The auxiliary-field Monte Carlo (AFMC) approach~\cite{Scalapino1980,Blankenbecler1981,Hirsch1985,Sugiyama1986}, also known in nuclear physics as the shell model Monte Carlo (SMMC) method~\cite{Lang1993,Alhassid1994,Koonin1997,Alhassid2001}, enables calculations in model spaces that are many orders of magnitude larger than those that can be treated by conventional diagonalization methods. AFMC is a powerful technique for calculating thermal and ground-state properties.  As a finite-temperature method,  it has been applied mainly to the calculation of statistical and collective properties of nuclei. In particular, AFMC is the state-of-the-art method for the microscopic calculation of nuclear level densities in the presence of correlations. (See Chap.~6 for the use of AFMC in  \emph{ab initio} applications  to light nuclei.)

While fermionic quantum Monte Carlo methods are often limited by the so-called sign problem that leads to large statistical errors, the dominant components~\cite{Dufour1996} of shell-model nuclear interactions have a good sign in AFMC and are often sufficient for realistic calculation of statistical and collective properties of nuclei. The smaller bad-sign components can be treated by the extrapolation method of Ref.~\cite{Alhassid1994}.

AFMC methods have been applied to other correlated quantum many-body systems. In condensed matter physics, AFMC has been used to study strongly correlated electron systems~\cite{Hirsch1985,Loh1992,Linden1992}. In quantum chemistry, it was applied to study the electronic structure of molecules, such as the recent study of the chromium dimer~\cite{Purwanto2015}.  In cold atom physics, AFMC methods were used to study the thermodynamics of the two-species Fermi gas with contact interaction for both the uniform gas~\cite{Bulgac2008,Bulgac2012} and the harmonically trapped gas~\cite{Gilbreth2013}, and the ground state of the Fermi gas in its unitary limit~\cite{Carlson2011}. AFMC simulations were recently carried out in studies of the neutron matter equation of state~\cite{Wlazlowski2014}.

Here we review the finite-temperature AFMC method in the context of the CI nuclear shell model, and in particular its recent developments and applications to mid-mass and heavy nuclei.  Earlier applications were discussed in Refs.~\cite{Koonin1997,Alhassid2001}  and references therein.  In Sec.~\ref{AFMC} we discuss the finite-temperature formalism of the AFMC method in the framework of both the grand canonical and canonical ensembles. The canonical ensemble, in which the number of particles is fixed, is particularly useful in applications to nuclei. In Sec.~\ref{projections} we discuss various projection methods in AFMC, and in Sec.~\ref{state-density} we describe the calculation of state densities. In Sec.~\ref{mid-mass} we discuss recent applications of AFMC to mid-mass nuclei, and in particular in the calculation of level densities, spin distributions and pairing gaps. In Sec.~\ref{heavy} we describe recent applications to heavy lanthanide nuclei, including the emergence of collectivity in the framework of the CI shell model, the calculation of state densities from the underlying Hamiltonian and the description of nuclear deformation in a rotationally invariant framework.  We conclude in Sec.~\ref{conclusion} with a summary and outlook. 

\section{Auxiliary-field Monte Carlo method}\label{AFMC}

The AFMC method is based on the Hubbard-Stratonovich (HS)~\cite{HS-trans,HS-transa} representation of the Gibbs ensemble, as we discuss in Sec.~\ref{sec-HS}.  The proper ensemble to describe nuclei is the canonical ensemble with fixed numbers of protons and neutrons, and in Sec.~\ref{particle-projection} we describe how this is accomplished using an exact particle-number projection~\cite{Ormand1994}. The Monte Carlo sampling method is briefly discussed in Sec.~\ref{MC}.  A rule to determine good-sign interactions in the grand canonical ensemble is discussed in Sec.~\ref{good-sign}. For a good-sign interaction, the Monte Carlo sign remains good when projecting on the canonical ensemble with even number of particles, but a sign problem emerges when projecting on an odd number of particles at low temperatures. In Sec.~\ref{odd-sign-prob} we describe a recent method we introduced to carry out accurate calculation of the ground-state energy of an odd-particle system despite the odd-particle sign problem~\cite{Mukherjee2012}.  

\subsection{The Hubbard-Stratonovich transformation}\label{sec-HS}

 The CI shell-model Hamiltonian $\hat H$ contains a one-body part described by single-particle orbitals $i$ and single-particle energies $\epsilon_i$ and a residual two-body interaction characterized by its two-body matrix elements $v_{ij,kl}$. The two-body interaction term can be brought to a diagonal form
\begin{eqnarray}\label{H}
\hat  H = \sum_i \epsilon_i \hat{n}_i
 +  \frac{1}{2} \sum_\alpha {v_\alpha} \hat{\rho}_\alpha^2  ,
 \end{eqnarray}
where $\hat{\rho}_\alpha$ are linear combinations of one-body densities  $\hat{\rho}_{ij} = a_i^\dagger a_j$,  and  $v_\alpha$ are the interaction ``eigenvalues.''

 The Gibbs density operator  $e^{-\beta H}$ at inverse temperature $\beta=1/kT$  can be viewed as the many-body evolution operator in
imaginary time $\beta$.  The HS transformation~\cite{HS-trans,HS-transa} expresses
 this propagator as a functional integral over one-body propagators describing non-interacting nucleons in
time-dependent external fields $\sigma(\tau)$ ($0 \leq \tau \leq \beta$).
The HS transformation is derived by dividing the time interval $(0,\beta)$ into $N_t$ time slices of length $\Delta \beta$ each, and factorizing
$e^{-\beta \hat H}  =  \left(e^{-\Delta \beta \hat H}\right)^{N_t}$.  For each time slice $\Delta\beta$, we have  to order  $(\Delta\beta)^2$ 
\begin{equation}\label{factorization}
  e^{-\Delta \beta \hat H} \approx  \prod_i e^{- \Delta \beta \epsilon_i \hat{n}_i} \prod_\alpha e^{-\frac{1}{2}{\Delta}\beta {v_\alpha}  \hat{\rho}_\alpha^2} .
 \end{equation}
 Each factor in the product over $\alpha$  in Eq.~(\ref{factorization}) can be written as an integral over an auxiliary variable $\sigma_\alpha$
\begin{eqnarray}
 e^{-\frac{1}{2} { \Delta}\beta {v_\alpha} \hat{\rho}_\alpha^2}
 = \sqrt{{\Delta\beta| v_\alpha|}\over{2\pi}} \int  d\sigma_\alpha
 e^{-\frac{1}{2} {\Delta}\beta |v_\alpha| {\sigma_\alpha^2}}
 e^{-{ \Delta}\beta |v_\alpha|  s_\alpha \sigma_\alpha \hat{\rho}_\alpha}
,
\end{eqnarray}
where $s_\alpha=\pm1$ for $v_\alpha < 0$ and $s_\alpha=\pm \mathrm{i}$ for $v_\alpha > 0$.  Using a set of auxiliary fields   $\sigma_\alpha(\tau_n)$ at each time
slice $\tau_n = n \Delta \beta$ and taking the limit of large $N_t$, we obtain the HS transformation 
\begin{eqnarray}\label{HS}
 e^{-\beta \hat H} = \int {\cal D}[\sigma] G_\sigma \hat U_\sigma ,
\end{eqnarray}
where
\begin{equation}
G_\sigma =  e^{-\frac{1}{2} \int_0^{\beta} |v_\alpha| \sigma_\alpha^2(\tau) d\tau }
\end{equation}
 is a Gaussian weight.  $\hat U_\sigma$ in Eq.~(\ref{HS}) is given by
\begin{equation}\label{prop}
\hat  U_\sigma ={\cal  T}  e^{-\int_0^\beta \hat h_\sigma(\tau) d\tau } ,
\end{equation}
where ${\cal T}$ denotes time ordering and
\begin{eqnarray}
\hat h_\sigma(\tau) = \sum_i \epsilon_i \hat{n}_i
 +\sum_\alpha s_\alpha |v_\alpha| \sigma_\alpha(\tau) \hat{\rho}_\alpha  
\end{eqnarray}
 is a one-body Hamiltonian describing nucleons moving in external time-dependent auxiliary fields $\sigma_\alpha(\tau)$.  The measure in the functional integral (\ref{HS}) over the auxiliary fields is 
\begin{eqnarray}
 {\cal D}[\sigma] \equiv \prod_{\alpha, n}\left[d\sigma_\alpha
(\tau_n)  \sqrt{{{\Delta}\beta |v_\alpha| /{2\pi}}}\right]
.
\end{eqnarray}

Using the HS transformation (\ref{HS}), the thermal expectation of an observable $\hat O$ can be written as
\begin{eqnarray}\label{observ}
\langle \hat O \rangle=
{{\rm Tr}\,( \hat O e^{-\beta \hat H})\over{\rm Tr}\,(e^{-\beta \hat H})} =
{\int {\cal D}[\sigma] G_\sigma \langle \hat O \rangle_\sigma{\rm Tr}\,\hat U_\sigma
\over \int {\cal D}[\sigma] G_\sigma {\rm Tr}\,\hat U_\sigma}  ,
\end{eqnarray}
 where  $\langle \hat O \rangle_\sigma\equiv
 {\rm Tr} \,( \hat O \hat U_\sigma)/ {\rm Tr}\,\hat U_\sigma$ is the expectation value of $\hat O$ for non-interacting particles in external auxiliary fields $\sigma(\tau)$.

\subsubsection{Algebraic structure of the Hubbard-Stratonovich transformation}\label{algebraic} 

The set of all one-body densities $a^\dagger_i a_j$ ($i,j=1,\ldots, N_s$, where $N_s$ is the number of single-particle orbitals) forms a Lie algebra, i.e., the commutator of any two such one-body densities is a linear combination of one-body densities. The corresponding operators of the form $\exp(\sum_{ij} c_{ij} a^\dagger_i a_j)$ (where $c_{ij}$ are c-numbers)  therefore describe a Lie group.  The one-body propagator $\hat U_\sigma$ in Eq.~(\ref{prop}) is a time-ordered product of such group elements and therefore is also a group element. The operator $\hat U_\sigma$ defines a propagator in Fock space for any number of particles. Of particular interest is its representation in the single-particle space, described by an $N_s\times N_s$  matrix ${\bf U}_\sigma$.   As we discuss below, the quantities appearing in the integrands of Eq.~(\ref{observ}) can be expressed in terms of this single-particle representation.  

We first discuss the grand canonical ensemble, in which the traces are evaluated over the many-particle Fock space with all possible particle numbers.  A chemical potential $\mu$ (which determines the average number of particles in the ensemble) is introduced by replacing the single-particle energies $\epsilon_i$ with $\epsilon_i - \mu$.  

\subsubsection{One-body observables} 

The trace of $\hat U_\sigma$  over the complete Fock space is given in terms of the matrix ${\bf U}_\sigma$  by
\begin{equation}\label{partition}
{\rm Tr}\; \hat U_\sigma = \det ( {\bf 1} + {\bf U}_\sigma) .
\end{equation}
Eq.~(\ref{partition}) can be thought of as the grand-canonical partition function of non-interacting fermions in a given set of time-dependent external fields $\sigma(\tau)$.

The grand canonical expectation value of a one-body operator $\hat O = \sum_{i,j} O_{ij} a^\dagger_i a_j$ can be calculated from
\begin{equation}\label{1-body}
\langle a_i^\dagger a_j \rangle_\sigma \equiv {{\rm Tr}\; (a_i^\dagger a_j  \hat U_\sigma) \over {\rm Tr}\; \hat U_\sigma } 
 = \left[ {1 \over {\bf 1} +{\bf U}^{-1}_\sigma }
\right]_{ji} .
\end{equation}

\subsubsection{Two-body observables}

Since $\hat U_\sigma$ describes an uncorrelated ensemble, we can use Wick's theorem to calculate the grand canonical expectation value of a two-body operator 
\begin{equation}
\langle a^\dagger_i a^\dagger_j a_l a_k \rangle_\sigma = \langle a^\dagger_i a_k \rangle_\sigma  \langle
 a^\dagger_j a_l\rangle_\sigma - \langle a^\dagger_i a_l \rangle_\sigma \langle
 a^\dagger_j a_k\rangle_\sigma ,
 \end{equation} 
  where the expectation values on the right-hand side are given by Eq.~(\ref{1-body}).

\subsection{Particle-number projection}\label{particle-projection}

 The nucleus is a finite-size system, and it is important to consider the canonical ensemble with fixed numbers of protons and neutrons.  We are therefore  interested in canonical expectation values.  Such quantities can be calculated using an exact particle-number projection.  Since the number $N_s$ of single-particle orbitals is finite, we can describe the particle-number projection by a discrete Fourier transform. The canonical partition function of $\hat U_\sigma$  for particle number ${\cal A}$  is given by~\cite{Ormand1994}
 \vspace{8pt}
\begin{eqnarray}\label{canonical}
{\rm Tr}_{\cal A} U_\sigma =\frac{e^{-\beta\mu {\cal A}}
}{N_s}\sum_{m=1}^{N_s}
e^{-\mathrm{i}\varphi_m {\cal A}}\det \left( {\bf 1}+e^{\mathrm{i}\varphi_m}e^{\beta\mu}{\bf U}_\sigma\right)
,
\end{eqnarray}
where $\varphi_m=2\pi m/N_s \;\; (m=1,\ldots,N_s)$ are quadrature points and $\mu$
is a chemical potential introduced to stabilize the numerical evaluation of the Fourier sum.  Similarly for a one-body observable $\hat O=\sum_{i,j} O_{i,j} a^\dagger_i a_j$
\begin{eqnarray}\label{1-body-canonical}
{\rm Tr}_{\cal A} \left(\hat O \hat U_\sigma \right) =\frac{e^{-\beta \mu {\cal A}}}{N_s} & \sum_{m=1}^{N_s}
e^{-\mathrm{i}\varphi_m{\cal A}} 
\det \left( {\bf 1}+e^{\mathrm{i}\varphi_m +\beta\mu}{\bf U}_\sigma\right) \\ \nonumber &
\times  {\rm tr}\;\left( {1 \over {\bf 1} +e^{-\mathrm{i}\varphi_m-\beta\mu}{\bf U}^{-1}_\sigma } {\bf O}
\right) ,
\end{eqnarray}
where ${\bf O}$ is the matrix with elements $O_{ij}$.
In the actual nuclear calculations we project on both neutron number $N$ and proton number $Z$.

\subsection{Monte Carlo sampling}\label{MC}

The integrands in Eq.~(\ref{observ}) are calculated by matrix algebra in the single-particle space [see, e.g., Eqs.~(\ref{partition}) and (\ref{1-body})].  However, the number of integration variables $\sigma_\alpha (\tau)$ is very large.
 For small but finite $\Delta \beta$,  this multi-dimensional integral can be evaluated exactly (up to a statistical error) by Monte Carlo methods.

In the applications of AFMC to nuclei we carry out the Monte Carlo sampling in the canonical ensemble. For a nucleus of  ${\cal A}$  nucleons we define the positive-definite weight function
\begin{equation}
 W_\sigma \equiv
  G_\sigma \vert {\rm Tr}_{\cal A} \;  \hat U_\sigma \vert .
\end{equation}
Next we define the $W$-weighted average of a quantity $X_\sigma$ that depends on the auxiliary field configuration $\sigma$ by 
\begin{equation}\label{ave_x}
\overline{X}_{\sigma} \equiv \frac {\int D[\sigma ] W_\sigma  X_{\sigma} \Phi_{\sigma}} { \int D[\sigma]  W_\sigma \Phi_{\sigma}},
\end{equation}
where
\begin{equation}\label{sign}
\Phi_\sigma\equiv {\rm Tr}_{\cal A} \; U_\sigma /\vert {\rm Tr}_{\cal A} \; U_\sigma \vert
\end{equation}
 is the Monte Carlo sign.  With this definition, the canonical thermal expectation of an observable $\hat O$ can be written as 
  \begin{equation}
 \langle \hat O \rangle =  \overline{{ {\rm Tr}_{\cal A} ( \hat O \hat U_\sigma) / {\rm Tr}_{\cal A} \hat U_\sigma}}  .
 \end{equation}

  In AFMC, a random walk is performed in the space of auxiliary fields $\sigma \equiv\{\sigma_\alpha(\tau_m)\}$
 that samples the $\sigma$-fields according to  the positive-definite distribution $W_\sigma$.  The average $\overline{X_\sigma}$  is then estimated from
\begin{equation}\label{X-average}
\overline{X_\sigma}  \approx  {\sum_k  X_{\sigma^{(k)}}  \Phi_{\sigma^{(k)}}  \over \sum_k  \Phi_{\sigma^{(k)}} },
\end{equation}
where $\sigma^{(k)}$ are $M$ uncorrelated samples.
The statistical error of $\overline{X_\sigma}$ can be estimated from the variance of the ``measurements'' $X_{\sigma^{(k)}}$. Though a standard random walk can be constructed by  the Metropolis algorithm, a modification based on Gaussian quadratures
 improves its efficiency~\cite{Dean1993}.

\subsection{Sign problem and good-sign interactions}\label{good-sign}

Assuming the Hamiltonian in Eq.~(\ref{H}) is time-reversal invariant, it can be rewritten in the form
\begin{eqnarray}\label{trev}
\hat H=\sum_i \epsilon_i\hat{n}_i+ {1\over2}\sum_\alpha \tilde v_\alpha \left({\rho}_\alpha \bar{\rho}_\alpha  +  \bar{\rho}_\alpha
{\rho}_\alpha\right),
\end{eqnarray}
where $\bar{\rho}_\alpha$ is the time-reverse density of ${\rho}_\alpha$ and $\tilde v_\alpha $ are real.  When all the interaction eigenvalues  $\tilde v_\alpha$ in the representation (\ref{trev}) are {\em negative}, the grand canonical one-body partition function ${\rm Tr}\,\hat U_\sigma$ is positive for any sample $\sigma$. Such interactions are known as good-sign interactions in AFMC. 

To prove the above sign rule, we consider the one-body Hamiltonian that appears in the HS decomposition for the ensemble described by the Hamiltonian (\ref{trev}) 
\begin{equation}\label{1body-t}
\hat h_\sigma =\sum_i  \epsilon_i \hat{n}_i  + \sum_\alpha \left(\tilde v_\alpha s_\alpha \sigma^{*}_\alpha {\rho}_\alpha  +  \tilde v_\alpha s_\alpha \sigma_\alpha \bar{\rho}_\alpha \right).
\end{equation}
 When all $\tilde v_\alpha < 0$,  $s_\alpha=1$ for all $\alpha$ and the one-body Hamiltonian (\ref{1body-t}) is invariant under time reversal, i.e.,  $\bar h_\sigma = h_\sigma$.  Since the spins of the single-particle states are half integers, it follows that the eigenstates of the propagator matrix $\bf{U}_\sigma$ appear in time-reversed pairs with complex conjugate
 eigenvalues $\{\lambda_i,\lambda^\ast_i\}$. The grand canonical partition of $\hat U_\sigma$  in Eq.~(\ref{partition}) can then be written as 
 \begin{equation}
 {\rm Tr}\;\hat U_\sigma= \prod_i |1 +\lambda_i|^2 ,
 \end{equation}
 and is positive for any configuration $\sigma$ of the auxiliary fields.
 
 The dominant collective components  of effective nuclear interactions are attractive~\cite{Dufour1996}, and in our calculations of  statistical and collective properties of nuclei discussed here we used good-sign interactions (see Sec.~\ref{mid-mass-interaction} and Sec.~\ref{heavy-interaction}).  Small bad-sign components of realistic effective nuclear interactions can be treated following the extrapolation method of Ref.~\cite{Alhassid1994}. 

\subsection{Circumventing the odd particle-number sign problem}\label{odd-sign-prob}

For a good-sign interaction, the particle-number projected partition ${\rm Tr}_{\cal A} \hat U_\sigma$ for an even number of particles ${\cal A}$ remains almost always positive and the particle-projected Monte Carlo sign is good for an even-even nucleus. However, for an odd number of particles, the projected partition can be negative for certain field configurations (${\rm Tr}\; \hat U_\sigma$ is always real for a good-sign interaction). The average sign $\langle\Phi_\sigma\rangle$ is then smaller than 1 and it decreases with increasing values of $\beta$. At low temperatures, this so-called odd particle-number sign problem becomes severe. Consequently, it is difficult to determine an accurate ground-state energy for odd-mass and odd-odd nuclei.   

We introduced a method~\cite{Mukherjee2012} to circumvent the odd particle-number sign problem by using the imaginary-time single-particle Green's functions of the even-particle system to determine an accurate ground-state energy of the odd-particle system.  

The CI shell-model Hamiltonian is rotationally invariant, and the single-particle orbitals $i = (\nu m)$ with $\nu=(n l j)$ are characterized by a principal quantum number $n$, orbital angular momentum $l$, total angular momentum $j$ and its $z$ projection $m$.  For any $\nu$, we define the scalar Green's function 
\begin{equation}\label{green}
G_{\nu}(\tau) = { {\rm Tr}_{\cal A} \left[~e^{-\beta \hat H} \mathcal{T} \sum_m a_{\nu m}(\tau) a^\dagger_{\nu m}(0)\right] \over  {\rm Tr}_{\cal A}~e^{-\beta \hat H}},
\end{equation} 
where $\mathcal{T}$ denotes time ordering and $a_{\nu m}(\tau)\equiv e^{\tau \hat H} a_{\nu m} e^{-\tau \hat H}$ is an annihilation operator of a particle at imaginary time $\tau$ ($-\beta \leq \tau\leq \beta$) in an orbital $i=(\nu m)$. Using the HS representation and the notation of Eq.~(\ref{ave_x}), we obtain 
\begin{equation}
G_{\nu}(\tau) = \left \{ \begin{array}{ll}
                \overline{\displaystyle \sum_{m}\left [{\bf U}_{\sigma}(\tau) ({\bf 1}- \langle\hat\rho \rangle_{\sigma} )\right ]_{\nu m, \nu m}} & \mbox{ for }\tau>0\\
                                    &   \\
                \overline{\displaystyle \sum_{m} \left [ \langle \hat\rho \rangle_{\sigma} {\bf U}^{-1}_{\sigma}(|\tau|) \right ]_{\nu m, \nu m}} & \mbox{ for }\tau \leq 0
                \end{array} \right . .
\end{equation}
 Here ${\bf 1}$ is the identity matrix in the single-particle space, and $\langle \hat\rho \rangle_{\sigma}$ is a matrix whose $\nu m,\nu' m'$ matrix element  is $\langle \hat\rho_{\nu m,\nu' m'} \rangle_{\sigma}$ where $\hat\rho_{\nu m,\nu'm'} = a^\dagger_{\nu' m'} a_{\nu m}$.

\begin{figure}[h]
\centerline{\includegraphics[width=\textwidth]{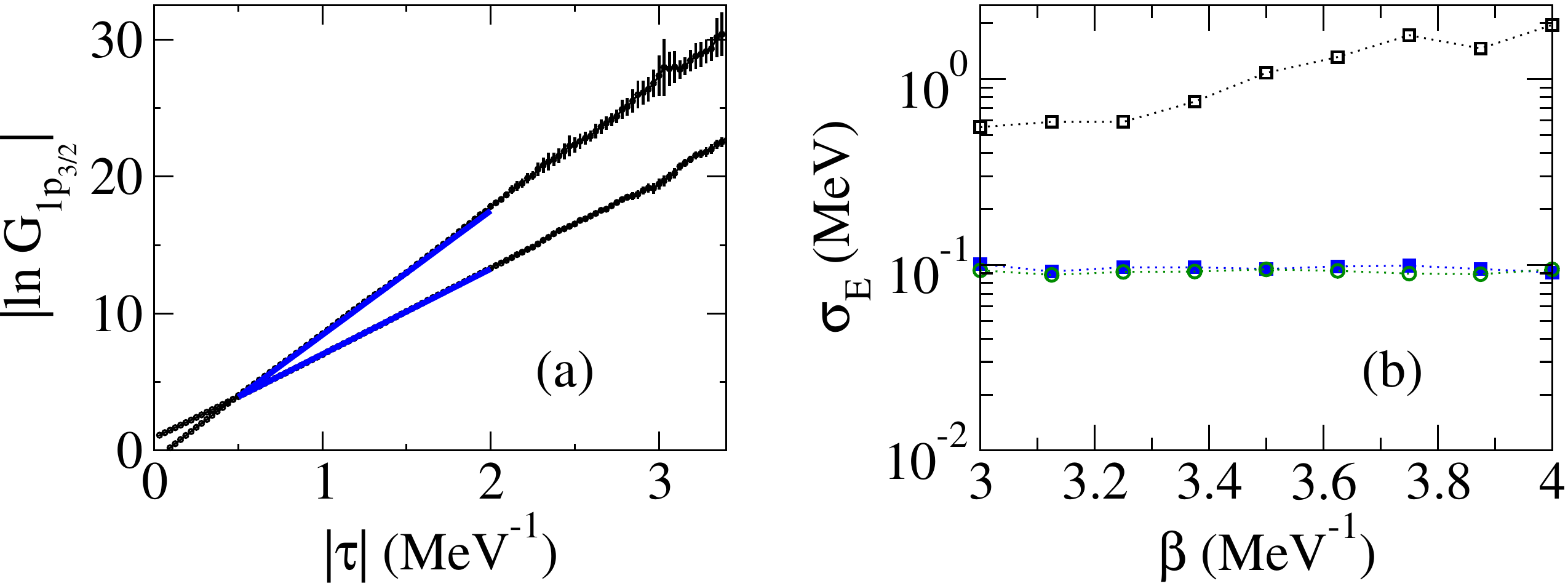}}
\caption{ (a)  Extracting the ground-state energy of the odd-even nucleus $^{57}$Fe in the imaginary time Green's function method.  The absolute value of logarithm of the AFMC Green's function for the neutron orbital $\nu=1p_{3/2}$ in $^{56}$Fe (lower curve) and $^{58}$Fe (upper curve) versus $|\tau|$ at $\beta=4$ MeV$^{-1}$.  The solid lines are linear fits for $0.5$ MeV$^{-1}\leq |\tau|\leq 2$ MeV$^{-1}$.
(b)  The statistical errors for the energy of $^{57}$Fe in the Green's function method (solid squares) are compared with statistical errors in direct AFMC calculations (open squares). A logarithmic scale is used for the statistical errors.  We also show for comparison the statistical errors associated with the energy of the even-even $^{56}$Fe  nucleus (open circles, almost on top of the solid squares). Adapted from Ref.~\cite{Mukherjee2012}. }
\label{odd-sign}
\end{figure}

Consider an even-even nucleus ${\cal A} \equiv (Z,N)$. Assuming that the ground state of this nucleus has zero spin, the Green's functions behave asymptotically in $\tau$  as  $G_{\nu}(\tau) \sim e^{- \Delta E_{J=j}({\cal A}_{\pm}) |\tau|}$ [${\cal A}_\pm$ denotes the even-odd nuclei $(Z,N \pm 1)$ when $\nu$ is a neutron orbital, and the odd-even nuclei $(Z \pm 1,N)$ when $\nu$ is a proton orbital].  The $+$ and $-$ correspond, respectively, to $\tau > 0$ and $\tau \leq 0$, and $\Delta E_{J=j}({\cal A}_{\pm})$ is the difference between the energies of the lowest spin $J$ eigenstate of the ${\cal A}_{\pm}$-particle nucleus and the ground state of the ${\cal A}$-particle nucleus.  In this asymptotic regime we determine $\Delta E_j({\cal A}_{\pm})$ from the slope of $\ln G_{\nu}(\tau)$. We then minimize $\Delta E_j({\cal A}_{\pm})$ over all possible values of $j$ to find the difference between the ground-state energy of the ${\cal A}_{\pm}$ nuclei  and the ground-state energy of the ${\cal A}$ nucleus, $E_{\rm gs}({\cal A})$.  Since $E_{\rm gs}({\cal A})$ and $G_{\nu}(\tau)$ characterize the even-even nucleus, they can be calculated  in AFMC without a sign problem (for a good-sign interaction).

We demonstrate the method for calculating the ground-state energy of $^{57}$Fe in Fig.~\ref{odd-sign} (the model space and interaction used are discussed in Sec.~\ref{mid-mass}). The left panel shows the absolute value of the logarithm of the Green's function for the neutron $1p_{3/2}$ orbital versus $|\tau |$ for the even-even nuclei $^{56}$Fe and $^{58}$Fe. The right panel shows that the statistical errors of the ground-state energy  of $^{57}$Fe extracted from the Green's function method  are much smaller than the statistical errors in direct AFMC calculations. 

\section{Projection methods}\label{projections}

The grand canonical traces we take in, e.g., Eq.~(\ref{observ}), are over the complete Fock space, in which case we can use Eqs.~(\ref{partition}) and (\ref{1-body}).   Canonical traces in the finite nucleus can be calculated via particle-number projection; see Eqs.~(\ref{canonical}) and (\ref{1-body-canonical}). 

The calculation of thermal observables at given values of good quantum numbers such as spin and parity, requires additional projections. Parity projection in AFMC was discussed in Refs.~\cite{Nakada1997,Nakada1998,Alhassid2000,Ozen2007}.  In Sec.~\ref{spin-projection} we discuss spin projection~\cite{Alhassid2007} in AFMC.  For isospin projection in AFMC see Ref.~\cite{Nakada2008}. 

It is also possible to project on observables that do not commute with the Hamiltonian. An example is the projection on the axial quadrupole operator~\cite{Alhassid2014} discussed in Sec.~\ref{q-projection}.

\subsection{Spin projection}\label{spin-projection} 

We first discuss the projection on the $z$ component $\hat J_z$ of the spin operator~\cite{Alhassid2007}. We define the projected partition function for an eigenvalue $M$ of $\hat J_z$ to be $Z_M(\beta) = {\rm Tr}_M e^{-\beta \hat H}$. Using the HS transformation and the notation of Eq.~(\ref{ave_x}), we have
\begin{equation}\label{M-ratio}
{Z_M(\beta) \over Z(\beta)} =  \overline{{{\rm Tr}_M \hat U_\sigma \over {\rm  Tr} \hat U_\sigma }}  .
\end{equation}
 The $M$-projected partition ${\rm Tr}_M \hat U_\sigma$ can be calculated using the Fourier representation of $\delta(\hat J_z - M)$ 
 \begin{equation}
  \label{M-project}
  {\rm Tr}_M \,\hat U_\sigma = {1 \over 2J_s + 1} \sum\limits_{k =-J_s}^{J_s}
  e^{-\mathrm{i} \varphi_k M} {\rm Tr}\,\left( e^{\mathrm{i}\varphi_k \hat J_z} \hat U_\sigma \right) ,
\end{equation}
where $\varphi_k= \pi k /(J_s+1/2)$ for $k=-J_s,\ldots,J_s$.
Since $\hat J_z$ is a one-body operator, we can use the group property (see Sec.~\ref{algebraic}) to represent $ e^{i\varphi_k \hat J_z} \hat U_\sigma$ in the single-particle space by the matrix $e^{i \varphi_k {\bf J}_z } {\bf U}_\sigma$ (${\bf J}_z$ is a diagonal $N_s\times N_s$  matrix with the magnetic quantum numbers of the corresponding orbitals along its diagonal). The grand canonical trace is then given by
\begin{equation}
 {\rm Tr}\,\left( e^{\mathrm{i}\varphi_k \hat J_z} \hat U_\sigma \right) =  \det \left( 1+e^{\mathrm{i} \varphi_k {\bf J}_z} {\bf U}_\sigma \right) .
\end{equation}
In practice, we also project on fixed numbers of protons and neutrons. 

The spin-projected partition function $Z_J(\beta) = {\rm Tr}_J e^{-\beta\hat H}$, is calculated using the identity ${\rm Tr}_J e^{-\beta \hat H} = {\rm Tr}_{M=J} e^{-\beta \hat H} - {\rm Tr}_{M=J+1} e^{-\beta \hat H}$ (valid since $ e^{-\beta \hat H}$ is a rotationally invariant operator). Using the HS transformation, we find
\begin{equation}
{Z_J(\beta) \over  Z(\beta)} = \overline{  {{\rm Tr}_{M=J} \hat U_\sigma \over {\rm  Tr} \hat U_\sigma } -  {{\rm Tr}_{M=J+1} \hat U_\sigma \over {\rm  Tr} \hat U_\sigma} } .
\end{equation}
We can similarly calculate the spin-projected energies, $E_M(\beta)$ and $E_J(\beta)$. 

\subsection{Axial quadrupole projection}\label{q-projection}

Here the observable of interest, the axial quadrupole operator $\hat Q_{20}= \sum_i \left(2  z_i^2 -  x_i^2 - y_i^2 \right)$, does not commute with the Hamiltonian, i.e., $[\hat H,\hat Q_{20}] \neq 0$. Its distribution $P_T(q)$ is given by~\cite{Alhassid2014}
\begin{equation}\label{prob1}
P_T(q) = \sum_n \delta(q - q_n) \sum_m \langle q,n | e,m\rangle^2 e^{-\beta e_m} ,
\end{equation}
where $|q,n\rangle$ are eigenstates of $\hat Q_{20}$ satisfying ${\hat Q_{20}}|q,n\rangle = q_n |q,n\rangle$ and similarly $|e, m\rangle$ are eigenstates of $\hat H$ with ${\hat H}|e,m\rangle = e_m |e,m\rangle$.  We note that in a finite model space the spectrum of $\hat Q_{20}$ is discrete.

$P_T(q)$ can be calculated in AFMC by a projection on $\hat Q_{20}$  using the Fourier transform of the Dirac $\delta$ function
\begin{equation} \label{prob}
P_T(q) \equiv { {\rm Tr}\, \left[\delta(\hat Q_{20} -q) e^{-\beta \hat H} \right] \over {\rm Tr}\, e^{-\beta \hat H} } = { \int_{-\infty}^\infty
{d \varphi \over 2 \pi} e^{-\mathrm{i} \varphi q }\, {\rm Tr}\, \left(e^{\mathrm{i} \varphi \hat Q_{20}} e^{-\beta \hat H} \right) \over {\rm Tr}\,  e^{-\beta \hat H}} 
\end{equation}
together with the HS representation (\ref{HS}) of $e^{-\beta \hat H}$.  In practice, we divide an interval $[-q_{\rm max},q_{\rm max}]$  into $2M+1$ intervals of length $\Delta q=2q_{\rm max}/(2M+1)$ and use a discrete Fourier representation for each sample $\sigma$
\begin{equation}\label{fourier-q}
{\rm Tr}\left[\delta(\hat Q_{20} - q_m) \hat U_\sigma \right] \!\!  \approx \!\! {1\over 2 q_{\rm max}}\! \! \sum_{k=-M}^M
  \!\!\!  e^{-\mathrm{i} \varphi_k q_m} {\rm Tr}(e^{\mathrm{i} \varphi_k \hat Q_{20}} \hat U_\sigma) ,
\end{equation}
where $q_m=m \Delta q$ ($m=-M,\ldots,M$) and $\varphi_k = \pi k/q_{\rm max}$ ($k=-M,\ldots, M$). Since $\hat Q_{20}$ is a one-body operator, we have
\begin{equation}
 {\rm Tr}(e^{\mathrm{i} \varphi_k \hat Q_{20}} \hat U_\sigma)  = \det \left({\bf 1} + e^{\mathrm{i} \varphi_k {\bf Q}_{20}} {\bf U}_\sigma \right) ,
 \end{equation}
 where ${\bf Q}_{20}$ is the matrix representing $\hat Q_{20}$ in the single-particle space. 
 
 \section{State densities}\label{state-density}
 
The nuclear state density is among the most important statistical nuclear properties. It is an integral part of the Hauser-Feshbach theory~\cite{Hauser1952} of statistical nuclear reactions and appears in the Fermi golden rule for transition rates.  However, the calculation of the state density in the presence of correlations is a challenging many-body problem, and most calculations are based on mean-field approximations such as the Hartree-Fock (HF) and the Hartree-Fock-Bogoliubov (HFB) approximations~\cite{Hilaire2006,Alhassid2015c}.  AFMC offers a state-of-the-art method for calculating state densities beyond the mean field in very large model spaces that are required at finite excitation energies~\cite{Nakada1997,Ormand1997,Langanke1998,Alhassid1999,Alhassid2003,Alhassid2008,Ozen2013}. 

In AFMC, we calculate the canonical thermal energy as the expectation value of the Hamiltonian, $E(\beta)=\langle H \rangle$. The canonical partition function $Z(\beta)$ can then be calculated by integrating the thermodynamic identity
 $-{\partial \ln Z / \partial \beta } = E(\beta)$. We find
\begin{equation}
\ln Z (\beta)= \ln Z (0) - \int_0^\beta E(\beta) d\beta ,
\end{equation}
where $Z(0)$ is the total number of many-particle states with $Z$ protons and $N$ neutrons in the model space. 
The state density $\rho(E)$ at energy $E$ is related to the partition function by an inverse Laplace transform
\begin{equation}\label{inv-L} 
\rho(E) = {1 \over 2\pi \mathrm{i}} \int_{-\mathrm{i} \infty}^{\mathrm{i} \infty} d\beta \,e^{\beta E} Z(\beta) . 
\end{equation}
The average state density is obtained by evaluating Eq.~(\ref{inv-L}) in the saddle-point approximation~\cite{BM1969} 
 \begin{equation}\label{saddle}
\rho(E) \approx \left( 2\pi T^2 C \right)^{-1/2}  e^{S(E)}, 
\end{equation} 
where $S(E)$ is the canonical entropy and $C$ is the canonical heat capacity given by 
\begin{equation}\label{S-C}
S =\ln Z + \beta E \;;\;\;\; C=\frac{dE}{dT} .
\end{equation}
The value of $\beta$ used in Eqs.~(\ref{saddle}) and (\ref{S-C})  is determined by the saddle-point condition 
$E = -{\partial \ln Z / \partial \beta } = E(\beta)$. 

Similar formulas apply for the calculation of the level density at a given value of a good quantum number by starting from the corresponding projected thermal energy as a function of $\beta$.  An example is the calculation of the spin-dependent level densities $\rho_J(E)$  by using the spin-projected energies $E_J(\beta)$ discussed in Sec.~\ref{spin-projection}.

\section{Applications to mid-mass nuclei}\label{mid-mass}

We first discuss applications of AFMC to mid-mass nuclei. In Sec.~\ref{mid-mass-interaction} we describe the model space and the interaction used. In Sec.~\ref{nickel} we discuss a recent calculation of level densities in a family of nickel isotopes. In Sec.~\ref{spin-distributions} we use the spin-projection method of Sec.~\ref{spin-projection} to calculate the spin distributions for several nuclei in the iron region, and in Sec.~\ref{pairing_gaps} we present results for pairing gaps in families of mid-mass isotopes calculated from odd-even mass differences. 

\subsection{Model space and interaction}\label{mid-mass-interaction}

We carried out AFMC studies of mid-mass nuclei using the $fpg_{9/2}$ shell for both protons and neutrons. The Hamiltonian used is an isoscalar. The single-particle energies are determined from a Woods-Saxon (WS) potential plus spin-orbit interaction with the parameters in Ref.~\cite{BM1969}.  The interaction is given by~\cite{Nakada1997}
\begin{equation}\label{H-fpg}
  -g \hat P^{(0,1)\dagger}\cdot \hat {\tilde P}^{(0,1)}
     - \chi \sum_\lambda k_\lambda \hat O^{(\lambda,0)}\cdot \hat O^{(\lambda,0)} ,
\end{equation}
where
\begin{eqnarray}\label{P-O}
 \hat P^{(\lambda,T)\dagger}&=&{\sqrt{4\pi}\over{2(2\lambda +1)}}
 \sum_{ab} \langle j_a\| Y_\lambda \|j_b\rangle  [a_{j_a}^\dagger \times a_{j_b}^\dagger]^{(\lambda,T)}, \nonumber\\
\hat O^{(\lambda,T)}&=&{1\over\sqrt{2\lambda +1}}  \sum_{ab} \langle j_a\| {{dV}\over{dr}} Y_\lambda \|j_b\rangle [a_{j_a}^\dagger \times \tilde a_{j_b}]^{(\lambda,T)} 
\end{eqnarray}
and $(\cdot)$ denotes a  scalar product in both spin ($\lambda $) and isospin ($T$). The modified annihilation operator $\tilde a$  is defined by $\tilde a_{j,m,m_t} = (-)^{j-m+{1\over 2}-m_t} a_{j,-m,-m_t}$, and  $\hat {\tilde P}^{(\lambda,T)}$ is similarly defined.
$V$ in Eq.~(\ref{P-O}) is the central part of the single-particle potential, and the three multipole interaction terms ($\lambda=2,3,4$)  are obtained 
by expanding the separable surface-peaked interaction $v({\bf r}, {\bf r}^\prime)  = -\chi (dV/dr)(dV/dr^\prime)  \delta(\hat{\bf r} - \hat{\bf r}^\prime)$.
The interaction strength $\chi$ is  determined by a self-consistency condition~\cite{Alhassid1996}.
Core polarization effects are taken into account by renormalizing $\chi$ with the factors $k_\lambda$.   We use $k_2=2$, $k_3=1.5$ and $k_4=1$,
in overall agreement with realistic effective nuclear interactions in this shell.
The pairing strength $g$ is determined so as to reproduce in particle-number projected BCS the pairing gaps determined from experimental odd-even mass differences for spherical  nuclei in the mass region $A=40$--$80$ with $Z=20$, $N=28$, $Z=28$ or $N=40$.  We determined a constant mean value of $g=0.212$~MeV.

\subsection{Level densities in nickel isotopes}\label{nickel} 

The density $\rho$ discussed in Sec.~\ref{state-density} is the {\em  state} density where the magnetic degeneracy $2J+1$ of levels with spin $J$ is included in the counting of states. However, the density measured in the experiments is often the {\em level} density $\tilde \rho$, where each level is counted just once, irrespective of its magnetic degeneracy.  In Ref.~\cite{Alhassid2015} we showed that it is possible to calculate directly the level density in AFMC by using
\begin{eqnarray}\label{level-density}
\tilde \rho =  \left\{ \begin{array}{cc} \rho_{M=0}   & \mbox{even-mass nucleus}  \\
\rho_{M=1/2}   &  \mbox{odd-mass nucleus} 
\end{array} \right. ,
\end{eqnarray}
where $\rho_M$ is the $\hat J_z$-projected density (see Sec.~\ref{spin-projection}). This density can be calculated as in Sec.~\ref{state-density} by replacing $E(\beta)$ with $E_M(\beta)$.

\begin{figure}[h]
\centerline{\includegraphics[width= \textwidth]{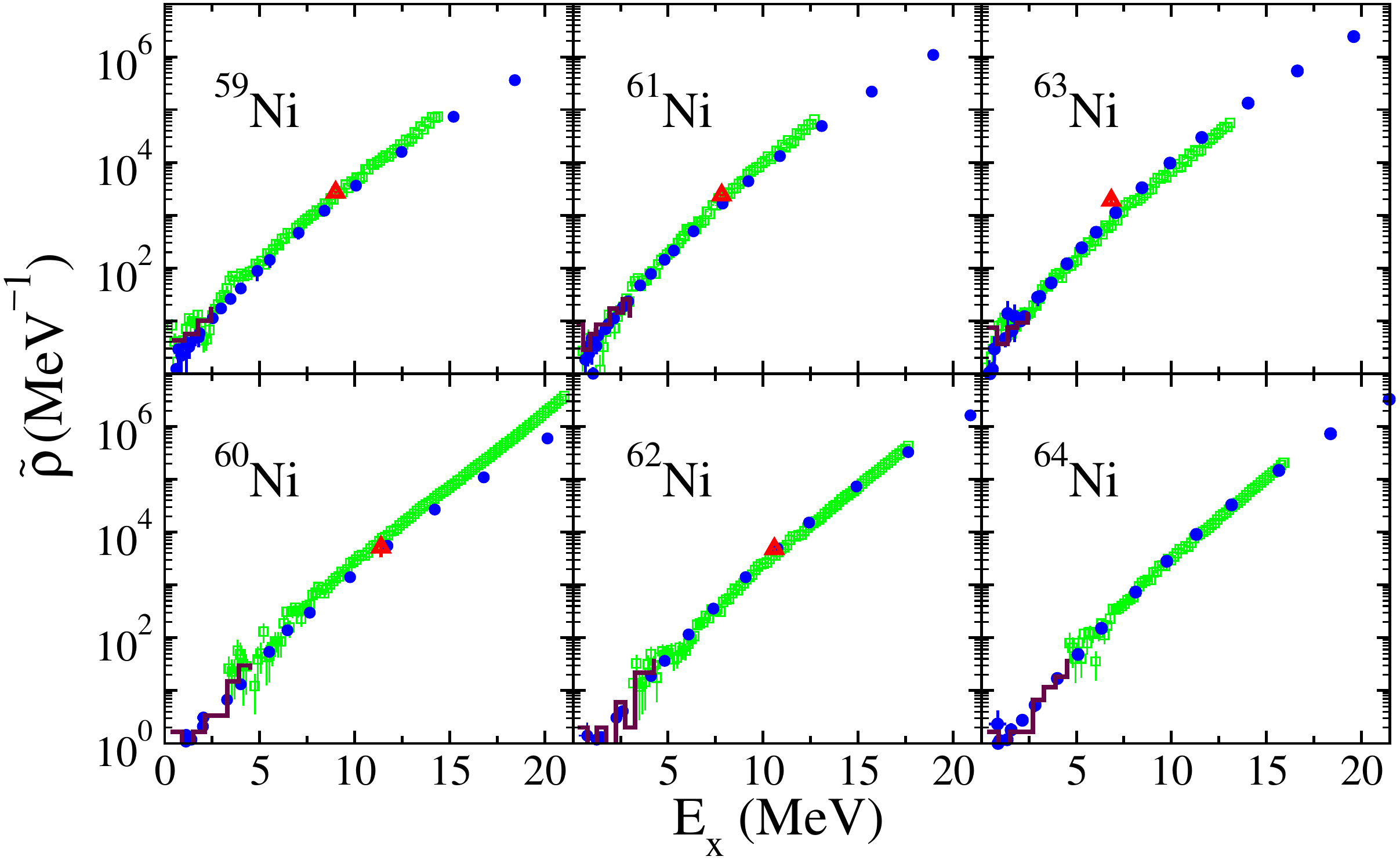}}
\caption{
Level densities $\tilde \rho$ in a family of nickel isotopes $^{59-64}$Ni: theory versus experiment.  The AFMC level densities (solid circles) are compared with level counting at low excitation energies (histograms), neutron resonance data (triangles)~\cite{Ripl3} when available, and level densities  extracted from proton evaporation spectra (squares forming quasi-continuous lines)~\cite{Voinov2012}.  Taken from Ref.~\cite{Bonett2013}.}
\label{Ni-level} 
\end{figure}

In Fig.~\ref{Ni-level} we show the AFMC level densities (solid circles) for a family of nickel isotopes $^{59-64}$Ni as a function of excitation energy~\cite{Bonett2013} calculated using Eq.~(\ref{level-density}).  Accurate ground-state energies for the odd-mass nickel isotopes  were determined by the Green's function method discussed in Sec.~\ref{odd-sign-prob}. The calculated level densities are in good agreement with experimental data: proton evaporation spectra (squares forming quasi-continuous lines)~\cite{Voinov2012}, level counting data (histograms) at low excitation energies and neutron resonance data (triangles)~\cite{Ripl3}  when available.

\subsection{Spin distributions}\label{spin-distributions}

We used the spin projection method of Sec.~\ref{spin-projection} to calculate the spin-projected energies $E_J(\beta)$ as a function of $\beta$. We can then use the saddle-point approximation of Sec.~\ref{state-density} to calculate  the spin-dependent level densities  $\rho_J(E_x)$ as a function of excitation energy $E_x$. 

In a statistical model in which the spins of individual nucleons are coupled randomly to total spin ${\bf J}$ the spin distribution $\rho_J/\rho$ follows the spin cutoff model~\cite{Ericson1960}
\begin{equation}\label{spin-cutoff}
 { \rho_J(E_x) \over \rho(E_x)} =  {(2J+1) \over 2\sqrt{2 \pi} \sigma^3} e^{-{J(J+1) \over 2  \sigma^2}},
\end{equation}
where the parameter $\sigma=\sigma(E_x)$ is known as the spin cutoff parameter. The spin-projected density $\rho_J(E_x)$ in Eq.~(\ref{spin-cutoff}) does not include the $2J+1$ magnetic degeneracy and is normalized by $\sum_J (2J+1) \rho_J(E_x)\approx \rho(E_x)$. The spin cutoff parameter is related to the thermal moment of inertia ${\cal I}$ (at temperature $T$) by 
\begin{equation}\label{inertia}
\sigma^2 = { {\cal I} T  \over \hbar^2} .
\end{equation}
In our AFMC studies of mid-mass nuclei we found that the spin cutoff formula (\ref{spin-cutoff}) works well at higher excitation energies with a rigid-body moment of inertia. However, in even-even nuclei we found~\cite{Alhassid2007} at low excitations an odd-even staggering effect in $J$. Also, we observed in such even-even nuclei a strong suppression of the moment of inertia at low excitations, an effect associated with pairing correlations. 

\begin{figure}[h]
\centerline{\includegraphics[width= \textwidth]{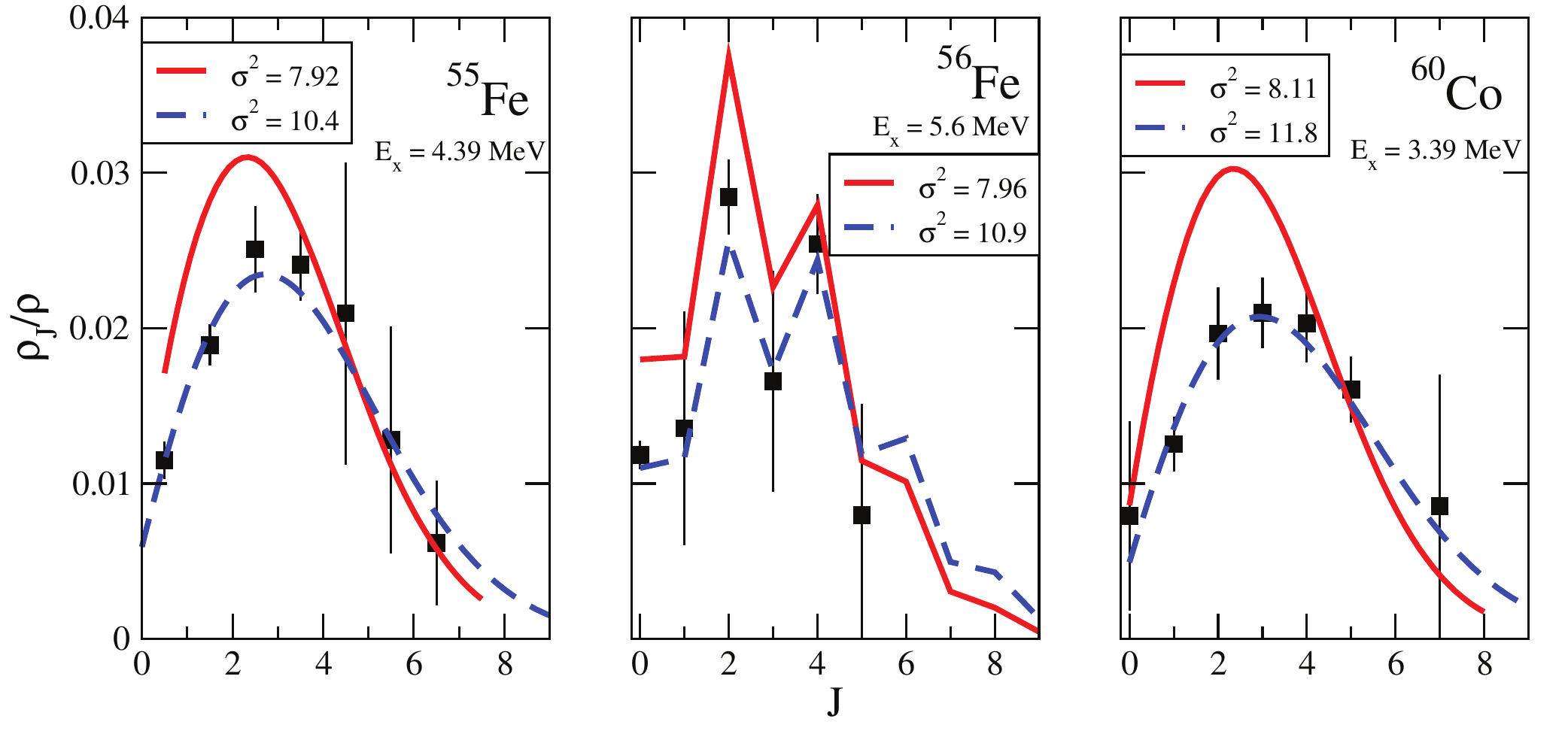}}
\caption{
Spin distributions $\rho_J/\rho$  as a function of spin $J$ for $^{55}$Fe (left panel),  $^{56}$Fe (middle panel) and $^{60}$Co (right panel) at excitation energies $E_x$. The AFMC distributions, taken from Ref.~\cite{Alhassid2007}, are shown by solid squares with statistical errors. The solid lines are empirical results~\cite{vonEgidy2008} based on systematic studies of low-lying experimental levels and the dashed lines describe similar distributions but with larger values of the spin cutoff parameter $\sigma$. Taken from Ref.~\cite{vonEgidy2008}.}
\label{spin-distribution} 
\end{figure}

In Fig.~\ref{spin-distribution} we show the spin distribution $\rho_J/\rho$ for the odd-even nucleus $^{55}$Fe, the even-even nucleus $^{56}$Fe and the odd-odd nucleus $^{60}$Co at the excitation energies specified in the figure.  The solid squares with statistical errors are the AFMC results of Ref.~\cite{Alhassid2007}.  The solid lines describe empirical distributions determined from experimentally known low-lying levels~\cite{vonEgidy2008,vonEgidy2009}.  The dashed lines describe similar empirical curves but with larger values of the spin cutoff parameter to account for the higher excitation energies used in the calculations. The empirical distributions are given by the spin cutoff model for the odd-even $^{55}$Fe and the odd-odd $^{60}$Co nuclei.  The staggering in the even-even $^{56}$Fe nucleus is described empirically by a spin cutoff formula multiplied by a factor $1 + x$ where $x \approx 0.23$ for even spin values, $x\approx -0.23$ for odd spin values, and $x \approx 1.02$ for $J=0$.  This empirical staggering is in good agreement with the AFMC predictions of Ref.~\cite{Alhassid2007}. 

\subsection{Pairing gaps}\label{pairing_gaps}

\begin{figure}[h]
\centerline{\includegraphics[width= \textwidth]{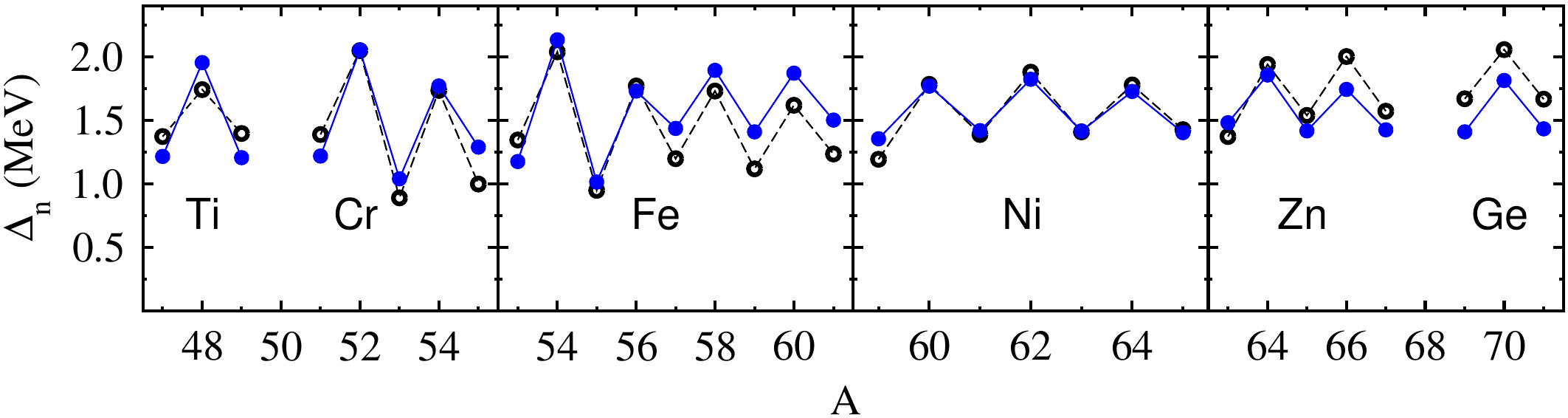} }
\caption{Neutron pairing gaps $\Delta_n$ from odd-even mass differences in families of isotopes in the iron region. The AFMC results (solid circles) are compared with experimental values (open circles).  Adapted from Ref.~\cite{Mukherjee2012}.}
\label{pairing-gaps}
\end{figure}

Pairing gaps can be calculated from odd-even mass differences~\cite{BM1969}. Using the method of Sec.~\ref{odd-sign-prob} to circumvent the odd-particle sign problem, we calculated ground-state energies of odd-mass nuclei. We can then find accurate neutron pairing gaps from the second-order difference in the ground-state energy as a function of the number of neutrons. In Fig.~\ref{pairing-gaps} we show neutron pairing gaps $\Delta_n$ in families of isotopes in the iron region. The AFMC gaps  are compared with gaps determined from the experiments. 

\section{Applications to heavy nuclei}\label{heavy}

Here we discuss recent applications of AFMC to heavy rare-earth nuclei. In Sec.~\ref{heavy-interaction} we present the model space and interaction used. In Sec.~\ref{collectivity} we demonstrate the emergence of various types of collectivity in heavy nuclei using the framework of the CI shell model. In Sec.~\ref{state-densities-heavy} we present results for the AFMC state densities in families of samarium and neodymium isotopes.  Finally, in Sec.~\ref{deformation} we discuss the finite-temperature distributions of the quadrupole deformation using the rotationally invariant framework of the CI shell model. 

\vspace{-4pt}
\subsection{Model space and interaction}\label{heavy-interaction}

The AFMC method was extended to heavy nuclei using a proton-neutron formalism, in which protons and neutrons can occupy different shells~\cite{Alhassid2008}. In particular, we applied the method to the lanthanides. The single-particle model space we used is composed of the orbitals $0g_{7/2},1d_{5/2},1d_{3/2,}2s_{1/2},0h_{11/2},1f_{7/2}$ for protons and
$0h_{11/2},0h_{9/2},1f_{7/2},1f_{5/2},2p_{3/2},2p_{1/2}, 0i_{13/2},1g_{9/2}$ for neutrons. The total number of single-particle states is 40  for protons and 66 for neutrons. The single-particle levels and wave functions are determined from a central WS potential plus spin-orbit interaction.  The interaction we used  is similar to the one used in \sref{mid-mass-interaction} and for the proton-neutron formalism it is given by
 \begin{equation}
\!\! - \!\!\!\!\sum_{\nu=p,n} g_\nu \hat P^\dagger_\nu \hat P_\nu
 - \!\!\sum_\lambda \chi_\lambda :(\hat O_{\lambda;p} + \hat O_{\lambda;n})\cdot
 (\hat O_{\lambda;p} + \hat O_{\lambda;n}): \;,
\label{SMint}
\end{equation}
where $\hat P^\dagger_\nu = \sum_{nljm}(-)^{j+m+l}a^\dagger_{\alpha jm;\nu} a^\dagger_{\alpha j-m;\nu}$ ($\nu=p,n$ and $\alpha=nl$) is the $J=0$ pair creation operator, $::$ denotes normal ordering, and $\hat O_{\lambda;\nu}=\frac{1}{\sqrt{2\lambda+1}}
\sum_{ab}\langle j_a||\frac{d V_{\rm WS}}{dr}Y_\lambda||j_b\rangle [a^\dagger_{\alpha j_a;\nu}\times \tilde{a}_{\alpha j_b;\nu}]^{(\lambda)}$
is a surface-peaked multipole operator [$\tilde{a}_{jm}=(-1)^{j-m}a_{j-m}$]. We include quadrupole, octupole and hexadecupole terms with corresponding strengths $\chi_\lambda=\chi k_\lambda$. The parameter $\chi$ is determined self-consistently~\cite{Alhassid1996} and
$k_\lambda$ are renormalization factors that account for core polarization effects. The values used for $g_\nu$ and $k_\lambda$ are given in Refs. \cite{Alhassid2008} and \cite{Ozen2013}.

The extension to heavy nuclei had required overcoming a number of technical challenges. A typical excitation energy in even-even rare-earth nuclei ($\sim 100$ keV) is an order of magnitude smaller than a typical excitation energy in even-even mid-mass nuclei.  It is then necessary to carry out the calculations to much lower temperatures to reach the ground-state energy. The larger values of $\beta$ and the larger band width of the single-particle spectrum require more computationally intensive calculations and lead to ill-conditioned matrices ${\bf U}_\sigma$.  Stabilization methods were introduced in strongly correlated electron systems in the grand canonical ensemble~\cite{Loh1992,Linden1992} and we extended them to the canonical ensemble~\cite{Alhassid2008}. Since there are $N_s$ terms in the Fourier sum of the canonical projection (see Sec.~\ref{particle-projection}), the computational cost scales as $N_s^4$. We recently introduced a novel stabilization method~\cite{Gilbreth2015} for the canonical ensemble that reduces this scaling to $N_s^3$.  The method was used in AFMC studies of cold atoms~\cite{Gilbreth2013} but it is also useful in the application of AFMC to nuclei.

\subsection{\mbox{Emergence~of~collectivity~in~the~configuration-interaction} shell-model approach}\label{collectivity} 
\begin{figure}[bth]
\centerline{\includegraphics[width=0.78\textwidth]{\figDIR 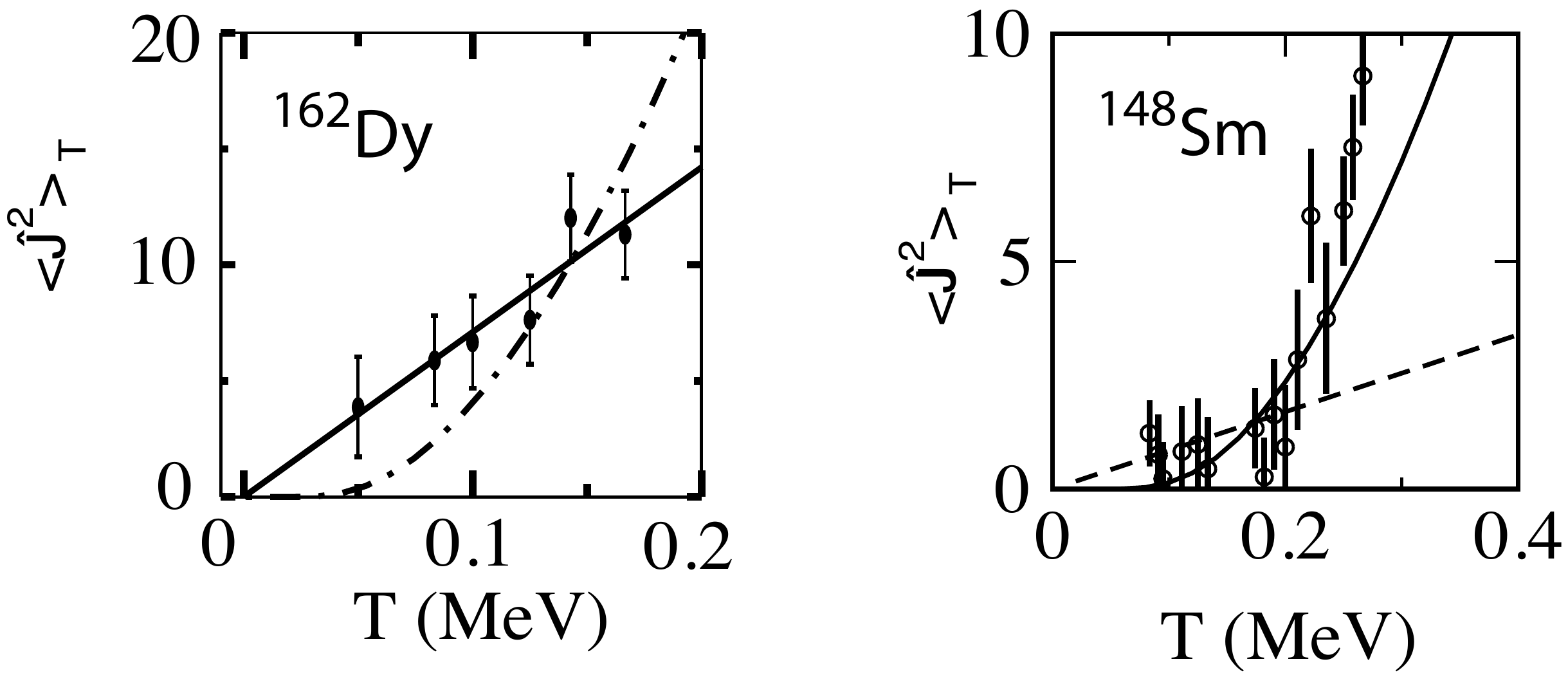}}
\caption{(Left) $\langle \mathbf{\hat J}^2 \rangle_T$ vs. temperature~$T$  in $^{162}$Dy.  The AFMC results (solid circles) are well described by a fit to the rotational band model (solid line). The dashed-dotted line is a fit to the vibrational model.  Adapted from Ref.~\cite{Alhassid2008}.   (Right) $\langle \mathbf{\hat J}^2 \rangle_T$ vs. temperature~$T$ in $^{148}$Sm. The SMMC results (open circles) are in better agreement with the vibrational model (solid line) than with the rotational model (dashed line). Taken from Ref.~\cite{Alhassid2015a}.}
\label{J2-vib-rot}      
\end{figure}

\vspace{-2pt}
Heavy nuclei are known to exhibit various types of collectivity that are well described by empirical models. However,  a microscopic description in the framework of the CI shell model has been mostly lacking.  A particularly important question is whether rotational collectivity, typical of deformed nuclei, can be described within a truncated spherical shell-model approach. Various types of collectivity are usually identified by their energy level schemes. However, while AFMC enables CI shell-model studies of heavy nuclei in very large model spaces, it is difficult to use for extracting detailed spectroscopic information. In Ref.~\cite{Alhassid2008} we identified a thermal observable, $\langle {\bf \hat J}^2\rangle_T$ (${\bf \hat J}$ is the total angular momentum of the nucleus), whose low-temperature behavior distinguishes between different types of collectivity.  Assuming an even-even nucleus with either a vibrational or rotational ground-state band, we find
\vspace{-5pt}
\begin{eqnarray}\label{J2-theory}
\langle \mathbf{\hat J}^2 \rangle_T \approx
 \left\{ \begin{array}{cc}
 30 { e^{-E_{2^+}/T} \over \left(1-e^{- E_{2^+}/T}\right)^2} & \mbox{vibrational band}  \\
 \frac{6}{E_{2^+}} T & \mbox{rotational band}
 \end{array} \right. ,
\end{eqnarray}
where $E_{2^+}$ is the excitation energy of the lowest $2^+$ level and $T$ is temperature.  In Fig.~\ref{J2-vib-rot} we show the low-temperature behavior of $\langle \mathbf{\hat J}^2 \rangle_T$ for $^{162}$Dy  (left panel) and $^{148}$Sm (right panel). The AFMC results (circles) are compared with the best fit (using the $2^+$ excitation energy as the fit parameter) to the formulas in (\ref{J2-theory}) for a nucleus with a rotational or vibrational band structure. We observe that in  $^{162}$Dy, the AFMC results for  $\langle {\bf \hat J}^2\rangle_T$  are in agreement with a straight line characterizing the rigid response to temperature in a rotational nucleus, while in $^{148}$Sm the AFMC results are in agreement with a softer response to temperature describing a vibrational nucleus.

\subsubsection{Crossover from vibrational to rotational collectivity} 

\begin{figure}[h]
\centerline{\includegraphics[width=\textwidth]{\figDIR 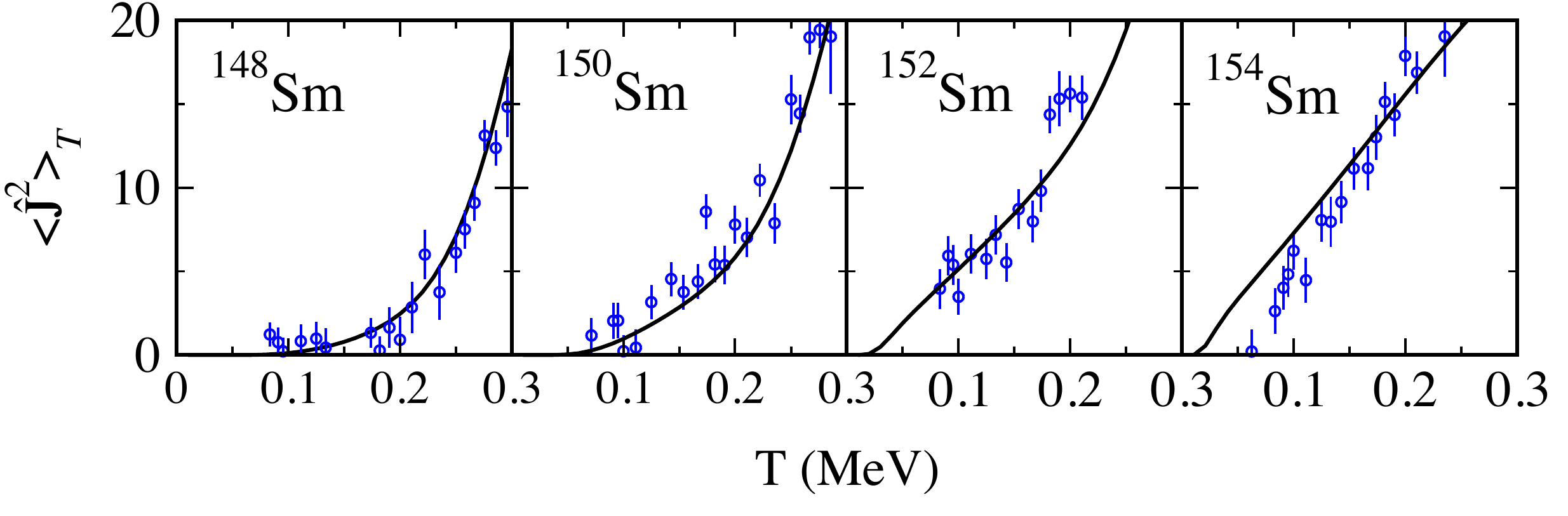}}
\caption{ $\langle {\bf \hat J}^2\rangle_T$ vs.~temperature $T$ for the even-mass samarium isotopes $^{148-154}$Sm. The AFMC results (open circles with statistical error bars) are compared with curves that are calculate from experimental data as discussed in the text  (solid lines). Adapted from Ref.~\cite{Ozen2013}.}
\label{Sm-J2}
\end{figure}

The observable $\langle \mathbf{\hat J}^2 \rangle_T$ can also be used to describe the crossover from vibrational to rotational collectivity. This is demonstrated in Fig.~\ref{Sm-J2} for a family of even-even samarium isotopes $^{148-154}$Sm. The AFMC results  for $\langle \mathbf{\hat J}^2 \rangle_T$ (open circles with statistical errors) are compared with values  deduced from the experiments (solid lines). We observe a crossover from a soft response to temperature in the vibrational nucleus $^{148}$Sm to a rigid response in the rotational nucleus $^{154}$Sm. 
The AFMC results are in overall agreement with the experimentally deduced curves. 
The latter are calculated from
\begin{eqnarray}\label{J2high}
 \langle \mathbf{\hat J}^2 \rangle_T  =  \frac{1}{Z(T)} \left(\sum_i^N J_i(J_i+1)(2J_i+1)e^{-E_{i}/T}  \right. \nonumber \\
    \left. + \int_{E_{N}}^\infty d E_x \: \rho(E_x) \: \langle \mathbf{\hat J}^2 \rangle_{E_x} \; e^{-E_x/T} \right),
\end{eqnarray}
where the partition function $Z(T)$ is 
\begin{equation}\label{Zhigh}
Z(T)=\sum_{i}^{N} (2J_i+1) e^{-E_i/T} + \int_{E_{N}}^\infty d E_x \rho(E_x) e^{-E_x/T} .
\end{equation} 
  The summations in Eqs.~(\ref{J2high}) and (\ref{Zhigh}) are over a set of experimental levels with excitation energies $E_i$ and spins $J_i$ that  is complete up to an energy $E_N$. Above $E_N$ the summation is replaced by an integral over a back-shifted Fermi gas formula whose parameters are determined empirically from level counting at low excitation energies and the neutron resonance data at the neutron separation energy. 

\subsection{State densities}\label{state-densities-heavy}  

\begin{figure}[h!]
\centerline{\includegraphics[width=\textwidth]{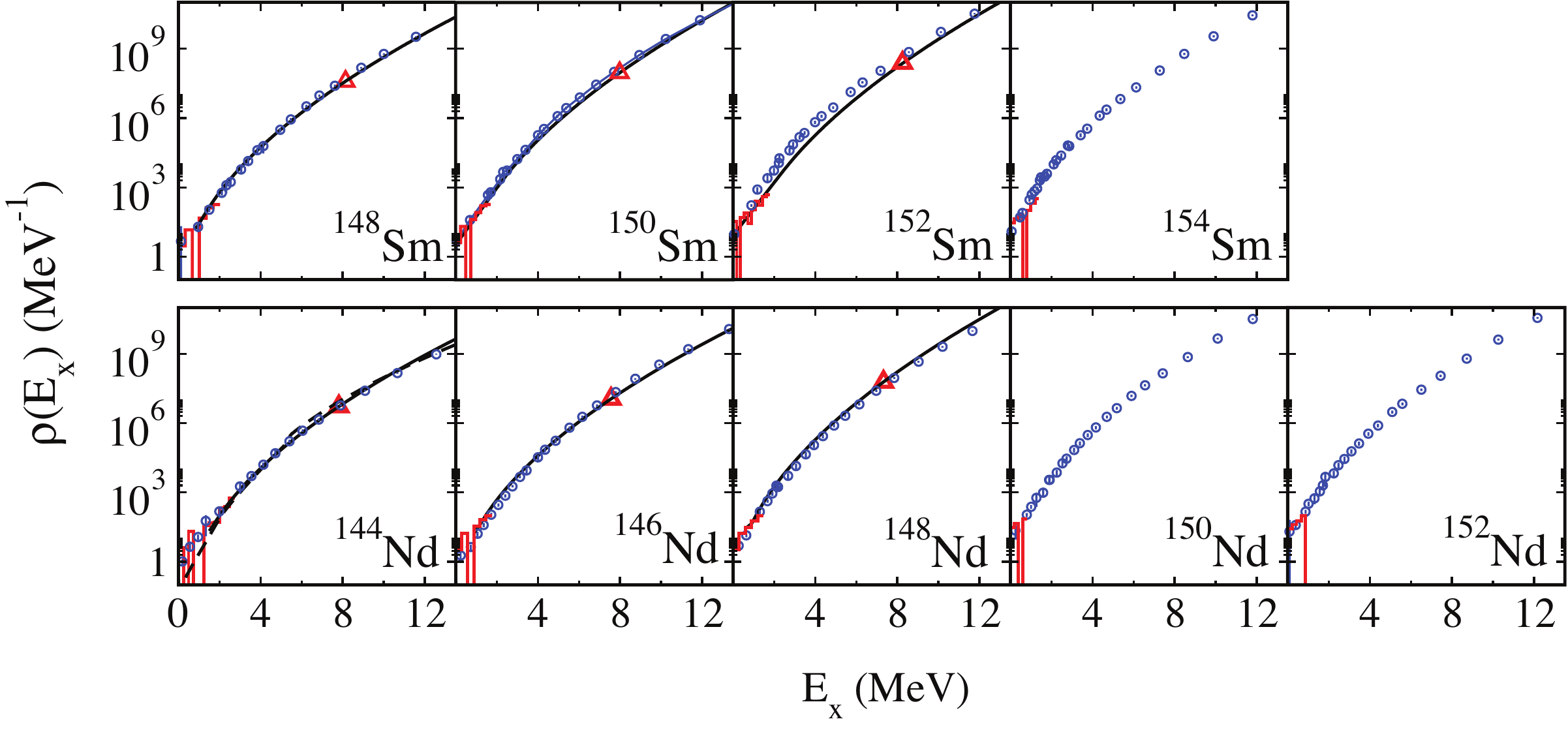}}
\caption{State densities of even-mass samarium isotopes (top row) and neodymium isotopes (bottom row) vs.~excitation energy $E_x$. The AFMC results (open circles) are overall in good agreement with level counting data at low excitation energies (histograms) and neutron resonance data (triangles) when available. The latter are converted to state densities assuming a spin cutoff formula (\ref{spin-cutoff}) with a rigid-body moment of inertia in Eq.~(\ref{inertia}). The solid lines describe empirical back-shifted Fermi gas formula with parameters determined from the experiments. 
 Adapted from Refs.~\cite{Ozen2013} and \cite{Alhassid2014a}.}
\label{rho_even}
\end{figure}

We calculated the state densities of families of even-even samarium and neodymium isotopes using the method of Sec.~\ref{state-density}. Fig.~\ref{rho_even} shows the AFMC state densities (open circles) of even mass $^{148-154}$Sm and $^{144-152}$Nd isotopes. The calculated densities are compared with level counting data at low excitation energies (histograms)  and neutron resonance data (triangles). 

\begin{figure}[h!]
\centerline{\includegraphics[width=0.85\textwidth]{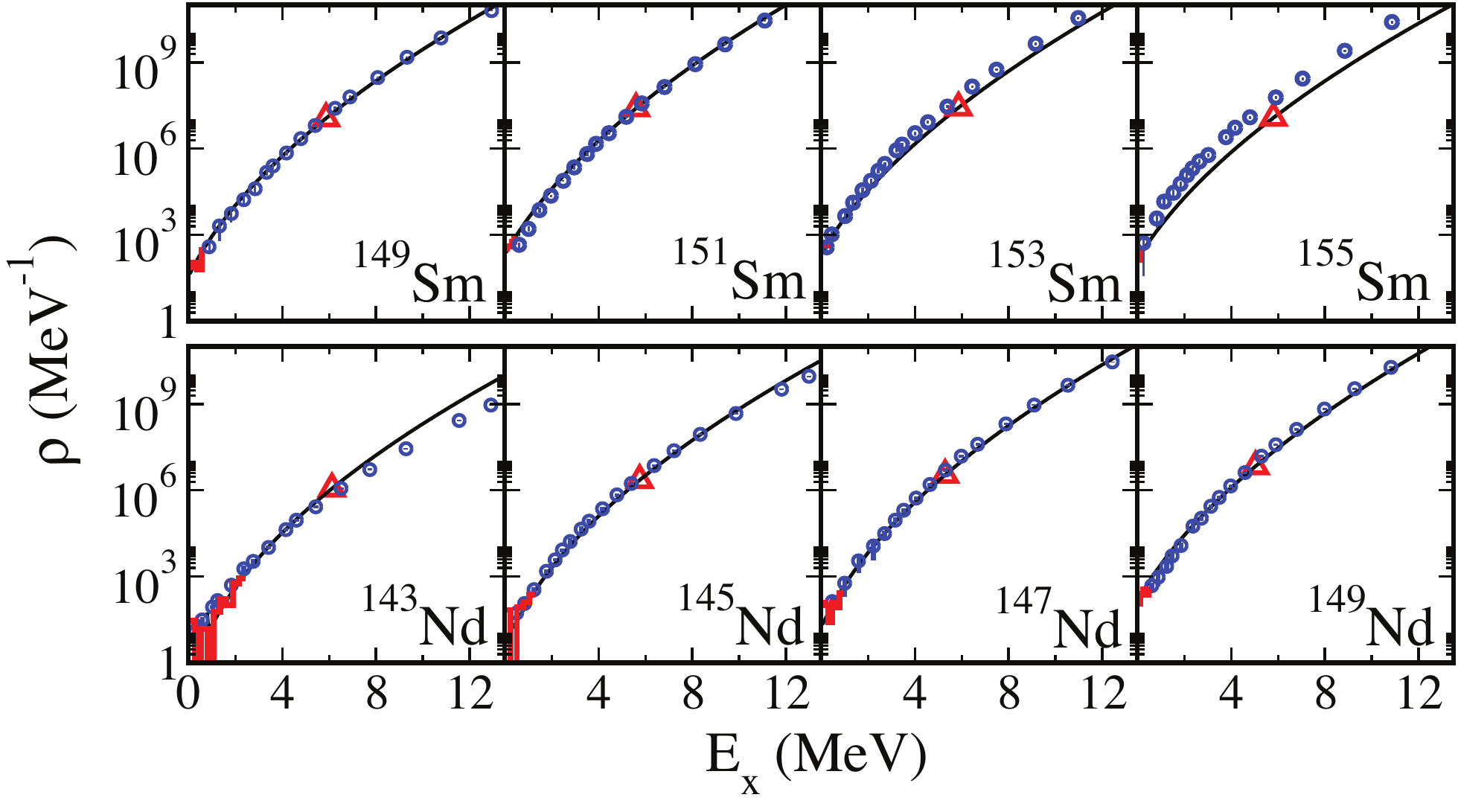}}
\caption{State densities vs.~excitation energy $E_x$ in odd-mass samarium (top row) and neodymium (bottom row) isotopes.  The AFMC densities are compared with experimental data. Symbols and lines are as in Fig.~\ref{rho_even}. Adapted from Ref.~\cite{Ozen2015}.}
\label{rho_odd}
\end{figure}

For odd-mass samarium and neodymium isotopes, AFMC calculations can be carried out in practice up to $\beta \sim 5$ MeV$^{-1}$.  At higher values of $\beta$, the statistical errors become too large because of the odd-particle sign problem. The method of Sec.~\ref{odd-sign-prob} is too time consuming for heavy nuclei and requires additional developments.  As an alternative we implemented a method~\cite{Ozen2015} to determine the ground-state energy of a heavy odd-mass nucleus by a one-parameter fit of its AFMC thermal energy  (at $T \geq 0.2 $ MeV for which AFMC results are available) to the thermal energy that is determined from experimental data. The results are shown in Fig.~\ref{rho_odd} for odd-mass samarium and neodymium isotopes.  We observe good agreement with experimental data. 

\subsection{\mbox{Nuclear deformation in a rotationally invariant framework}}\label{deformation}

Nuclear deformation is an important concept in understanding the structure of heavy nuclei. However, it is introduced in the context of a mean-field approximation (e.g., Hartree-Fock, HF, or Hartree-Fock-Bogoliubov, HFB), in which the rotational invariance is broken.  An important question is whether we can observe model-independent signatures of deformation within a framework that preserves rotational symmetry and still calculate an intrinsic deformation in such a framework. 

\subsubsection{Quadrupole distributions in the laboratory frame}

\begin{figure}[htb]
\centerline{\includegraphics[width=\textwidth]{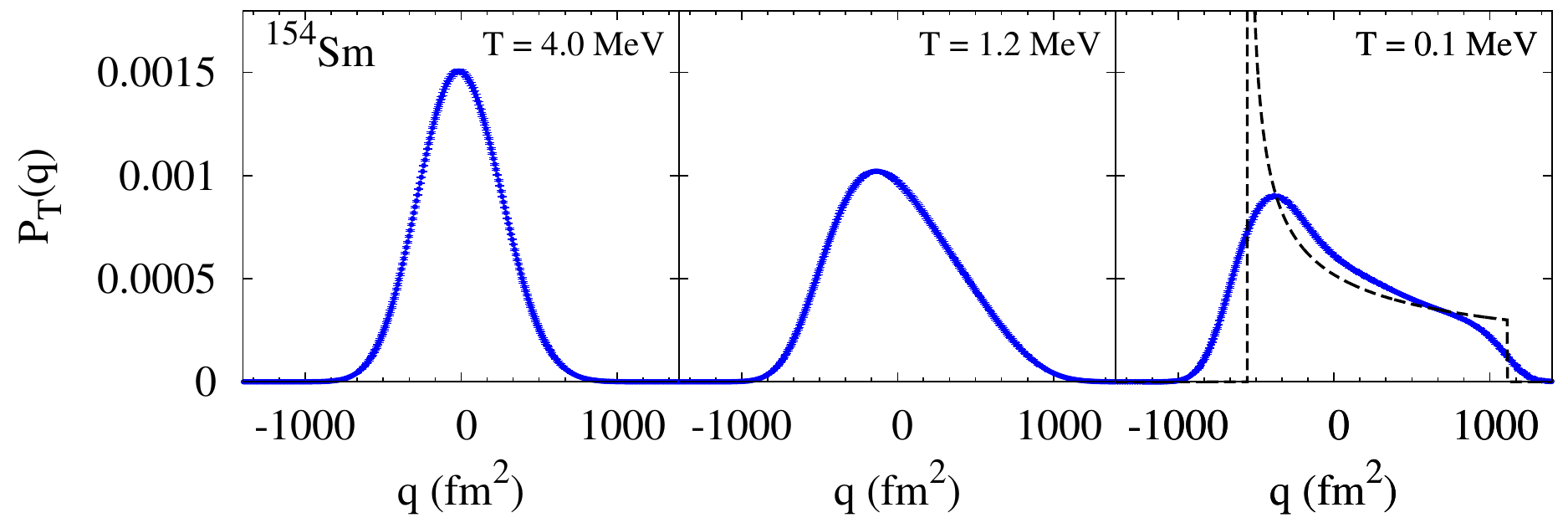}}
\caption{Axial quadrupole distributions $P_T(q)$ (see text) in $^{154}$Sm at temperatures $T=0.1$ MeV, $T=1.2$ MeV (close to the HFB shape transition temperature) and $T=4$ MeV. The dashed line at $T=0.1$ MeV is the corresponding quadrupole distribution for a prolate rotor with the same intrinsic quadrupole moment as that found in HFB.  Adapted from Ref.~\cite{Alhassid2014}.}
\label{Sm154_zvq}
\end{figure}

We used the quadrupole-projection method of Sec.~\ref{q-projection} to determine the distribution $P_T(q)$ of the axial quadrupole operator $\hat Q_{20}$ [see Eq.~(\ref{prob})] in the laboratory frame for heavy rare-earth nuclei. In Fig.~\ref{Sm154_zvq} we show the distributions $P_T(q)$  for $^{154}$Sm at three temperatures. In a mean-field approximation, such as the HFB approximation, this nucleus is deformed in its ground state (i.e., at $T=0$) and it undergoes a shape transition to a spherical nucleus at a certain critical temperature. At a low temperature ($T=0.1$ MeV) we find that $P_T(q)$ is a skewed distribution, and it is in qualitative agreement with the axial quadrupole distribution of a prolate rigid rotor with an intrinsic quadrupole moment that is the same as the ground-state value found in HFB (dashed line in Fig.~\ref{Sm154_zvq}). The distribution $P_T(q)$ remains skewed at the shape transition temperature around $T=1.2$ MeV, suggesting that deformation effects survive beyond the shape transition temperature.  At a high temperature of $T=4$ MeV, we observe a symmetric distribution $P_T(q)$ that is close to a Gaussian. 

In contrast, similar calculations for $^{148}$Sm result in Gaussian-like distributions already at low temperatures.
 This is consistent with  $^{148}$Sm being spherical in its ground state within the HFB approximation. We conclude that the distribution $P_T(q)$ can be used as model-independent signature of deformation in a framework that preserves rotational symmetry. 

\subsubsection{Quadrupole distributions in the intrinsic frame}

The quadrupole distribution $P_T(q)$ calculated in AFMC describes the laboratory-frame distribution. However, in modeling dynamical  nuclear processes, such as fission, it is often of interest to determine the statistical nuclear properties as a function of the intrinsic deformation. The intrinsic frame is a concept introduced in the context of a mean-field approximation and the challenge is to describe intrinsic deformation in the context of the rotationally invariant CI shell-model framework without resorting to a mean-field approximation. 

The distribution of the quadrupole tensor components $q_{2\mu}$ ($\mu=-2,\ldots,2$) is invariant under rotations. We can therefore expand the logarithm of the distribution in the so-called quadrupole invariants~\cite{Kumar1972,Cline1986}. There are three such invariants up to fourth order, which in terms of the intrinsic quadrupole deformation parameters\footnote{
We use $\beta$ to denote both the inverse temperature and the deformation parameter. The meaning should be clear from the context. 
},
$\beta$ and $\gamma$, are given by $\beta^2, \beta^3\cos 3\gamma$ and $\beta^4$.  Expanding to fourth order, we have 
\begin{equation}\label{landau}
-\ln P_T(\beta,\gamma) = N + A\beta^2 -B \beta^3\cos 3\gamma + C \beta^4 + \ldots ,
\end{equation}
where $A,B, C$ are temperature-dependent parameters and $N$ is a normalization constant. Eq.~(\ref{landau}) resembles the Landau expansion of the free energy in which the quadrupole tensor is treated as the order parameter of the shape transition~\cite{Levit1984,Alhassid1986,Alhassid1992, Alhassid1992a}. We can determine the parameters $A,B,C$ in Eq.~(\ref{landau}) from the expectation values of the three lowest-order quadrupole invariants. These expectation values can be calculated as a function of $A,B,C$  using the density distribution $P_T$ in Eq.~(\ref{landau}) together with the corresponding volume element  
\begin{equation}
\prod_\mu d q_{2\mu} = \frac{1}{2}  \beta^4 |\sin 3\gamma | \, d\beta \, d\gamma\, d\Omega ,
\end{equation}
where $\Omega$ are the Euler angles characterizing the orientation of the intrinsic frame. On the other hand, the expectation values of the three lowest-order quadrupole invariants are related to moments of $\hat Q_{20}$ in the laboratory frame~\cite{Alhassid2014} 
\begin{equation}
\langle \hat Q \cdot \hat Q \rangle = 5 \langle \hat Q_{20}^2 \rangle\;;\; \langle (\hat Q \times \hat Q) \cdot \hat Q \rangle = -5 \sqrt{\frac{7}{2}} \langle \hat Q_{20}^3 \rangle \;;\; \langle (\hat Q \cdot \hat Q)^2 \rangle = \frac{35}{3} \langle \hat Q_{20}^4 \rangle 
\end{equation}
and can therefore be directly calculated from the AFMC distribution $P_T(q)$. 

\begin{figure}[htb]
\centerline{\includegraphics[width=\textwidth]{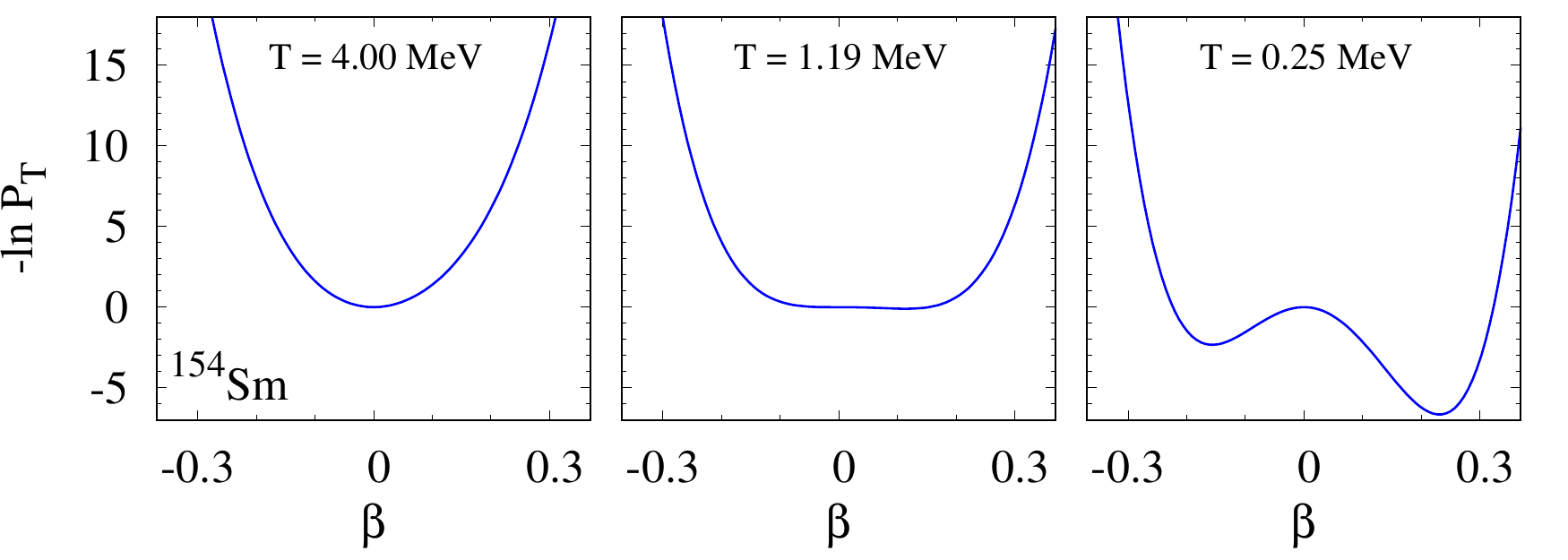}}
\caption{$-\ln P_T(\beta,\gamma=0)$ in Eq. (\ref{landau}) versus quadrupole deformation parameter  $\beta$  in $^{154}$Sm at  temperatures $T=0.25, 1.19$ and $4$ MeV.  Taken from Ref.~\cite{Alhassid2015b}.}
\label{Sm154_logP}
\end{figure}

In Fig.~\ref{Sm154_logP} we show the calculated curves, $-\ln P_T(\beta,\gamma=0)$,  versus deformation $\beta$  for $^{154}$Sm at three temperatures.  Even though these curves are calculated in the CI shell-model approach,  which preserves rotational invariance, they seem to mimic the behavior of the free energy surfaces within a mean-field approximation describing a shape transition from a deformed prolate nucleus to a spherical nucleus.

The distributions $P_T(\beta,\gamma)$ at constant temperature $T$ can be converted to 
level densities $\rho(E_x,\beta,\gamma)$ as a function of excitation energy $E_x$ and intrinsic deformation $\beta,\gamma$ using the saddle-point approximation. 

\section{Conclusion and outlook}\label{conclusion}

The AFMC method is a powerful technique for calculating thermal and ground-state properties of many-particle fermionic systems in very large model spaces. It has been applied to strongly correlated electron systems, molecules, cold atomic Fermi gases and nuclei. Here we reviewed recent developments and applications of AFMC to nuclei in the framework of the nuclear CI shell-model approach. Of particular importance in nuclear applications of AFMC is the use of the canonical ensemble with fixed numbers of protons and neutrons. AFMC is a state-of-the-art method for calculating statistical properties of nuclei and,  in particular, level densities and their dependence on good quantum numbers such as spin and parity.  We also discussed the use of AFMC to calculate collective nuclear properties, e.g., pairing gaps and quadrupole deformation. We presented applications to mid-mass nuclei in the iron region and to heavy lanthanide nuclei.  

Interesting future AFMC studies in nuclei include the calculation of statistical nuclear properties as a function of the intrinsic quadrupole deformation, and the extension of the method to other mass regions in the table of nuclei, such as the actinides and unstable nuclei. 
The CI shell-model Hamiltonians are specific to each mass region. It would be useful to derive effective CI shell-model Hamiltonians from a theory that is valid globally across the table of nuclei, such as density functional theory. First steps in mapping an energy density functional onto a shell-model Hamiltonian were discussed in Refs.~\cite{Alhassid2006} and \cite{Guzman2008}.  

\section*{Acknowledgments}

This work was supported in part by the Department of Energy grant No.\ DE-FG-0291-ER-40608.
I  would like to thank  G.F. Bertsch, M. Bonett-Matiz, L. Fang, C.N. Gilbreth, S. Liu, A. Mukherjee, H. Nakada, and C. \"Ozen  for their collaboration on various parts of the work reviewed above. The research presented here used resources of the National Energy Research Scientific Computing Center, which is supported by the Office of Science of the U.S. Department of Energy under Contract No.~DE-AC02-05CH11231.  It also used resources provided by the facilities of the Yale University Faculty of Arts and Sciences High Performance Computing Center.

\end{document}